# VST – VLT SURVEY TELESCOPE INTEGRATION STATUS

C. BELFIORE, M. BRESCIA, M. CAPACCIOLI, O. CAPUTI,
G. CASTIELLO, F. CORTECCHIA, L. FERRAGINA, D. FIERRO,
V. FIUME, D. MANCINI, G. MANCINI, G. MARRA,
L. MARTY, G. MAZZOLA, L. PARISI, L. PELLONE,
F. PERROTTA, V. PORZIO, P. SCHIPANI, G. SCIARRETTA,
G. SPIRITO, M. VALENTINO

*INAF – Osservatorio Astronomico di Capodimonte, Via Moiariello 16, I-80131 Napoli (Italy)*

G. SEDMAK

*Università degli Studi di Trieste, Dipartimento di Astronomia, Via G.B.Tiepolo 11, I-34131 Trieste ( Italy)*

The VLT Survey Telescope (VST) project started in 1997 at OAC with the aim of promoting the design and construction of a 2.6 m aperture, wide-field (1°×1°), UV to I facility, to be installed and operated at the European Southern Observatory (ESO) on the Cerro Paranal Chile. VST was primarily intended to complement the observing capabilities of VLT with wide-angle imaging for detecting and pre-characterising sources for further observations with the VLT.But the VST was also thought for non-VLT related stand-alone survey projects. The paper will report actual project development status during integration phase in progress.

## 1. Project Description

The VST (VLT Survey Telescope) program is a co-operation between OAC (Osservatorio Astronomico di Capodimonte) and ESO (European Southern Observatory) regulated by a Memorandum of Understanding approved by the ESO Council on June 1998. It is a 2.6m Alt-Az telescope to be installed at Cerro Paranal in Chile, in the ESO site, the same area as the VLT (Very Large Telescope).

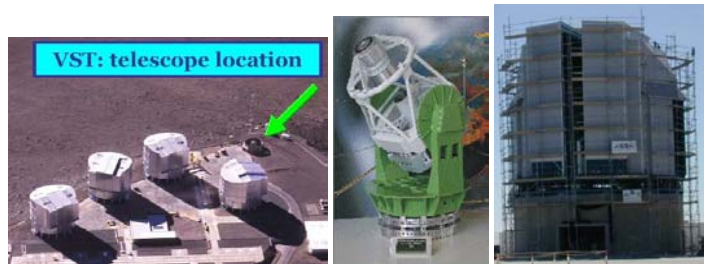

Figure 1 – VST location, model and building

The VST is a wide-field imaging facility planned to supply databases for the ESO VLT science and carry out stand-alone observations in the UV to I spectral range. For the telescope focal plane instrument, in January 2000 ESO signs a MoU with the OmegaCAM Consortium (ESO, Germany, Italy, Netherland) for the realization of a wide-field camera for VST, with a mosaic detector composed by 32 2Kx4K pixel size CCD.

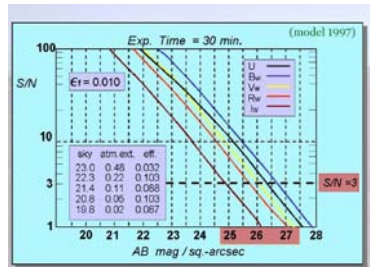

Figure 2 – VST + Ωcam expected performances

When used in combination with the VLT units it will have the potential of allowing astronomers to reach the frontier in ground-based optical astronomy at very high spatial resolution. For this reason the VST design has been carefully done in order to produce sharp and stable images. In addition, careful attention, paid to the telescope thermal control in combination with active optics, allow improving the angular image size beyond the performance of even larger but older telescopes. The VST design strategy has been defined in order to guarantee the optimization of the whole system and its technological sub-components. Also the software, controlling and handling the telescope operations, has been carefully designed following same strategy. The VST telescope plays a role of a powerful survey instrument for the largest ground-based telescope of the world, the ESO VLT. OAC has been committed to design and install the telescope, software included. After the installation and commissioning, VST will be managed by ESO staff. Therefore in order to simplify the future maintenance of the software by ESO, the easiest way is to develop the VST software in the most "VLT-compliant" way.

## 2. VST control system highlights

In the VST a distributed control driven through 4 CAN buses is adopted for:

 M1 mirror axial and radial pads control system
 Axes motor cooling and temperature acquisition systems
 Telescope structure temperature monitoring system
 M2 mirror control system

The VST AO is based on a complex control scheme; both primary and secondary mirrors must be supported and accurately positioned. The aberrations corrected by VST AO control system are: coma, defocus (M2 re-alignment), spherical, astigmatism, quad-astigmatism. VST has 84 axial pads distributed on four concentric rings, which include 12, 18, 24 and 30 pads. 24 lateral pads provide the active lateral supports balancing M1 during altitude axis motions.

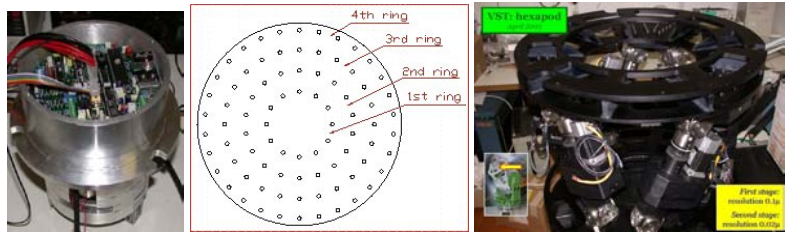

Figure 3 – from left: M1 axial pad control, M1 axial pad distribution and M2 hexapode

M1 is controlled by a set of active supports based on astatic levers; they are driven in order to elastically compensate wave-front deformations measured by a SH wavefront sensor in a feedback loop.

## 3. Integration Status

Actually the telescope is under integration phase. Current status is the following. 100% of mechanical parts available with 85% of them already mounted in the assembly area at MECSUD factory (45 km far from Napoli). Altitude motors already mounted on telescope arms. Main axes electronics, drive system and 80% of control software completed and currently under test at telescope. New M1 mirror under construction. M2 mirror available. M1 actuators at 80% ready. M2 active control hexapode already available at OAC. NEXT MILESTONE: telescope completely integrated and fully tested (including control software) in the first half of next year, (acceptance in Europe).